\documentstyle[11pt,newpasp,twoside,psfig]{article}
\index{summary}
\index{instructions}
\index{template}

\newcommand\xte{{\it RXTE}}

\newcommand\asca{{\it ASCA}}

\def\sp{\hskip 1.5pt}

\def\kes73{\hbox{Kes\sp73}}

\def\edcomment#1{\iffalse\marginpar{\raggedright\sl#1\/}\else\relax\fi}
\reversemarginpar

\begin{document}
\title{Hiccups in the night: X-ray monitoring of the two Crab-like LMC pulsars}

\author{E. V. Gotthelf }
\affil{Columbia Astrophysics Laboratory, 550 West 120th St, New York, NY 10027, USA}
\author{W. Zhang, F. E. Marshall} 
\affil{Laboratory for High Energy Astrophysics, Goddard Space Flight Center, Greenbelt, MD 20771}
\author{J. Middleditch}
\affil{Los Alamos National Laboratory, MS B256, CCS-3, Los Alamos, NM 87545}
\author{Q. D. Wang}
\affil{Department of Astronomy, University of Massachusetts, B-524, LGRT, Amherst, MA 01003}

\begin{abstract}

We are undertaking an extensive X-ray monitoring campaign of the two
Crab-like pulsars in the Large Magellanic Clouds, PSR~B0540-69 and PSR
J0537-6910. We present our current phase-connected timing analysis
derived from a set of 50 pointed X-ray observations spanning several
years. From our initial 1.2 yr monitoring program of the young 50 ms
pulsar PSR~B0540-69, we find the first compelling evidence for a
glitch in its rotation. This glitch is characterized by $\Delta \nu /
\nu = (1.90 \pm 0.05) \times 10^{-9}$ and $\Delta \dot \nu / \dot \nu
= (8.5 \pm 0.5) \times 10^{-5}$.  Taking into account the glitch
event, we derive a braking index of $n = 1.81 \pm 0.07$, significantly
lower than previous reported. For the 16 ms pulsar, PSR~J0537-6910, we
recorded 6 large glitch events during a period of nearly 3 years, the
highest rate of all known Crab-like systems.  Despite the extreme
timing activity, the long term spin-down of this pulsar continues to
average $-1.9743 \times 10^{-10}$ Hz/s.

\end{abstract}

\section{Introduction}

A characteristic signature of young rotation-powered pulsars is the
phenomena of ``glitches'', sudden discontinuous changes in their spin
periods (e.g., see Lyne \& Graham-Smith 1998). The physical causes of
these glitches are not understood. Suggestions include
sudden changes in the neutron star (NS) crust configuration
(``starquakes''), abrupt reconfiguration of the magnetic field, or
perhaps to the sudden unpinning of vortices in the superfluid neutrons
in the inner part of the NS crust. For the latter, the amplitude of the
glitch provides an estimate of the fractional part of the moment of
inertia carried by superfluid neutrons (Lyne et al. 1996).

The largest glitches have relative amplitudes ($\Delta\nu/\nu$) of
several parts per million, but the range of amplitudes covers many
orders of magnitude. Often there is a partial recovery back toward the
pre-glitch rotation rate on a time scale of $\sim 100$ days, however,
the spin-down rate may be permanently altered. Lyne (1995) noted that
the amount of recovery in the rotation rate tends to be inversely
proportional to the characteristic age of the pulsar.  For some
pulsars (e.g. Crab) the glitches are accompanied by a persistent
increase in the spin-down rate with a relative amplitude of a few
$\times 10^{-4}$. This increase may be caused, for example, by changes
in the alignment of the magnetic field because of starquakes (Allen \&
Horvath 1998; Link et al. 1998). 

In this paper we present preliminary results on the first detection of
glitches from the two LMC Crab-like pulsars.

\section{Observations and Results}

All data presented herein were obtained with the PCA instrument
on-board the Rossi X-ray Timing Explorer (\xte) for the purpose of
monitoring the 16 ms pulsar PSR~J0537-6910 (Marshall et
al. 1998). However, the 50 ms PSR~B0540-69, located $22'$ away, lies
well within the RXTE $1^{\circ}$ (FWHM) field of view and was thus
simultaneously observed. The duration and spacing between observations
were optimized to phase-link the pulses from PSR~J0537-6910 which is
more then sufficient to allow a similar analysis for PSR~B0540-69,
which is both slower and brighter than PSR~J0537-6910.

The PCA is sensitive to X-rays in the $2-60$ keV band, with a spectral
resolution $\delta E/E \sim 18\%$ at 6 keV and a time resolution of $<
100 \mu s$. To optimize the pulsar detection, we selected $\sim 2-15$
keV photons from the top layer of the PCA only. Photon arrival times were
corrected to the solar system barycenter.

Our timing analysis procedure is summarized below (see Zhang et
al. 2001 for details). For each observation we folded the arrival
times into 20 phase bins over a range of frequencies centered on an
initial value found from an FFT of the data set (without accounting
for any $\dot \nu$). From the folded profile we determined a pulse
time-of-arrival (TOA) for each observation in an iterative manner,
initially using a linear fit to the TOA's then adding higher order
derivative terms as needed.

\begin{figure}[t]
\small
      \begin{minipage}[t]{0.48\linewidth}
\psfig{file=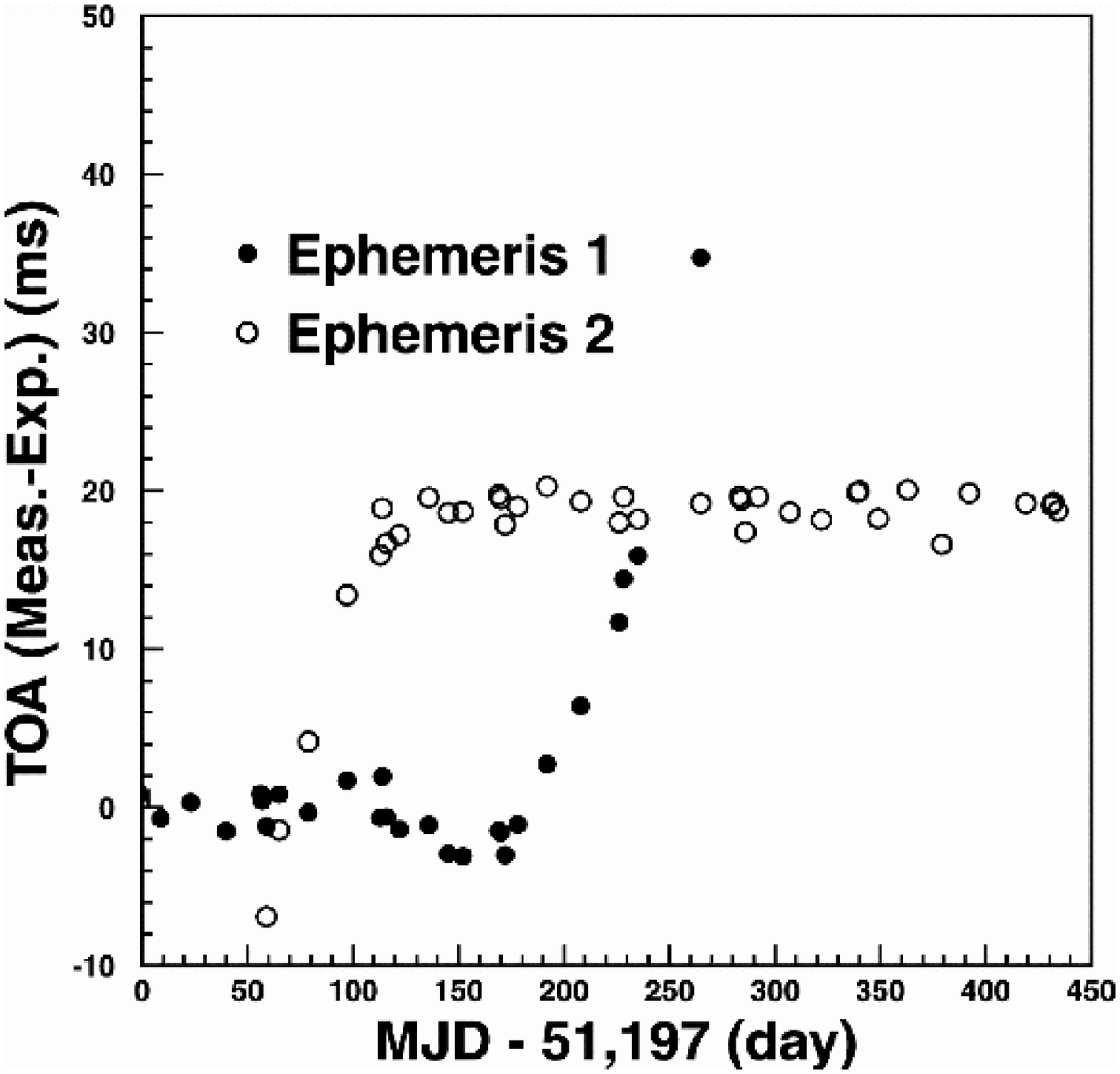,height=2.0in,angle=270.0} 
{
{\bf Figure 1.} Phase residuals vs. time for PSR~0540-69. The two sets of
symbols represent the pulse arrival times computed using two
ephemerides, one before and one after the glitch at MJD 51,325 
(Zhang et al. 2001).
}
      \end{minipage}\hfill
      \begin{minipage}[t]{0.48\linewidth}
\psfig{file=gotthelfe3_boston_fig2.ps,height=2.0in,angle=270.0} 
{
{\bf Figure 2.} Timing residuals for PSR~J0537-6910 based on the
monitoring observations presented herein. The data has been fit with
the linear ephemeris shown on the plot.
}
      \end{minipage}
\end{figure}

\subsection{\bf The 50~ms LMC Pulsar PSR~0540-69}

Using the above prescription we fitted the TOA's for the PSR~0540-69
data sets.  A linear fit was used as a quadratic term was not needed
to improve the fit. However, we found an apparent timing glitch had
occurred, too small to show up directly in a frequency residuals but
large enough to be evident in the analysis of the TOAs. An improved
fit was obtained by breaking up the data set up into two groups and
fitting each separately.  Figure 1 presents the pulse arrival times
fitted with the ephemeries of Zhang et al. (2001). We determine that
the glitch occurred at MJD $51,325 \pm 45$. The glitch at this epoch
changed in frequency and frequency derivative $\Delta \nu / \nu =
(1.90 \pm 0.05) \times 10^{-9}$ and $\Delta \dot \nu / \dot \nu = (8.5
\pm 0.5) \times 10^{-5}$, respectively; the magnitude of $\Delta \dot
\nu $ for this glitch is $\sim 3.5$ times larger than the one Deeter,
Nagase, \& Boynton (1999) suggested.

     Our timing result for PSR~B0540-69 is most consistent with the
extrapolated ephemeris of Deeter et al. (1987 to 1991) and agrees to
within the uncertainties, while that of Boyd (spanning 1979 to 1993)
over-predicts our measured frequencies by tens of micro-hertz. Our
braking index of $1.81 \pm 0.07$ is significantly lower than than the
values of $2.28 \pm 0.02$ (Boyd et al. 1995) and $2.0799 \pm 0.00027$
(Deeter et al. 1999).  Eikenberry et al. (1998) found $n = 2.5$ whose
errors are dominated by timing noise. However, all previous
measurements explicitly assumed that the pulsar had not suffered any
glitches during the observation intervals. 

Given our new result, and
the apparent inconsistent in the reported braking index, it is likely
that there may have been many other glitches in the timing history of
PSR~B0540-69. In particular, when we combine the data points in Table
1 of Boyd et al. (1995) with the frequency measurements of the work
presented herein, we obtain a braking index of $2.103 \pm 0.005$,
significantly lower than $2.28 \pm 0.02$. Based on the glitch rates
from other Crab-like pulsars, such as PSR~J537-6910 (see below) and on
Vela (nine glitches in 25 yr; Lyne et al. 1996), it is certainly
possible that PSR~B0540-69 could have an average glitch rate of one
per year. Since our measurements come from phase-linked data sets,
they likely provide the most accurate measure of the secular braking index.

\subsection{The 16~ms LMC Pulsar PSR~0537-6910}

Following a similar analysis as described above, for PSR~J0537-6910 we
find six large glitches occurred during the first 900 days of our
monitoring campaign. The intervals between glitches ranged between
$\sim 80$ to $\sim 300$ days. The relative increases of the pulse
frequency ranged from $\sim 0.14 \times 10^{-6}$ to $\sim 0.68 \times
10^{-6}$, and relative amplitude ranged from $0.15 - 0.68 \times
10^{-6}$ in frequency. Glitches with amplitudes as small as $0.01
\times 10^{-6}$ could have easily been detected.  Their absence
suggests that small glitches are relatively rare in PSR~J0537-6910.

With the exception of the Jan 2000 observations, we are able to
maintain the cycle-count of the pulse period for all data between the
glitches. The long term spindown of the pulsar continues to average
$-1.9743 \times 10^{-10}$ Hz/s, i.e., the same value previously
reported. With continued monitoring we should be able to determine the
braking index, $n$ for PSR~J0537-6910; this may allow us to
distinguish between possible physical mechanisms contributing to the
spindown.

Although PSR~J0537-6910 is the oldest and most rapidly rotating
Crab-like pulsar, the observed rate of glitching appears to be the
highest of all known Crab-like systems. The frequency and size of the
glitches are comparable to the extreme values for any pulsar.  The
pulsars exhibiting the largest known rates of glitches are
PSR~J1740-3015, for which nine glitches have been observed in 8 years
(Shemar \& Lyne 1996), and PSR~J1341-6220, for which 12 glitches have
been seen in 8.2 years (Wang et al. 2000).  Most of these glitches
were relatively small, however, with typical relative amplitudes of
$<1 \times 10^{-7}$.  However, the Vela pulsar has had nine major
glitches in 25 years of comparable size as those found for
PSR~J0537-6910 (Lyne et. al 1996).
   
\acknowledgments

This research is supported by the NASA LTSA grant NAG5-22250. J. M.
thanks the University of California's Institute for Nuclear and
Particle Astrophysics and Cosmology (INPAC) for funding.

\end{document}